\documentclass[twocolumn,superscriptaddress,preprintnumbers,amsmath,10pt,aps,prl,nobibnotes,nofootinbib]{revtex4-1} 

\usepackage{algorithm, algorithmic}

\usepackage{amsmath}
\usepackage{amsfonts}
\usepackage{bm}
\usepackage{color}
\usepackage{verbatim}
\usepackage{graphicx}
\usepackage{lipsum}
\usepackage[utf8]{inputenc}
\usepackage{times}
\usepackage{graphicx}
\usepackage{bm}
\usepackage{amssymb,multirow}
\usepackage[dvipsnames]{xcolor}
\usepackage[normalem]{ulem} 
\usepackage[mathlines]{lineno}
\usepackage{bbm}
\usepackage{siunitx}

\usepackage[utf8]{inputenc}
\usepackage[english]{babel}
\usepackage{amsmath}
\usepackage{enumitem}
\usepackage{graphicx}
\usepackage{amssymb}
\usepackage{float}
\usepackage{tikz}
\usepackage{moreenum}
\usepackage{gensymb}
\usepackage{braket}
\usepackage{mathrsfs}
\usepackage{xcolor}
\usepackage{cancel}

\usepackage{stackengine}
\stackMath

\makeatletter
\newlength\xvec@height%
\newlength\xvec@depth%
\newlength\xvec@width%
\newcommand{\xvec}[2][]{%
  \ifmmode%
    \settoheight{\xvec@height}{$#2$}%
    \settodepth{\xvec@depth}{$#2$}%
    \settowidth{\xvec@width}{$#2$}%
  \else%
    \settoheight{\xvec@height}{#2}%
    \settodepth{\xvec@depth}{#2}%
    \settowidth{\xvec@width}{#2}%
  \fi%
  \def\xvec@arg{#1}%
  \def\xvec@dd{:}%
  \def\xvec@d{.}%
  \raisebox{.2ex}{\raisebox{\xvec@height}{\rlap{%
    \kern.05em
    \begin{tikzpicture}[scale=1]
    \pgfsetroundcap
    \draw (.05em,0)--(\xvec@width-.05em,0);
    \draw (\xvec@width-.05em,0)--(\xvec@width-.15em, .075em);
    \draw (\xvec@width-.05em,0)--(\xvec@width-.15em,-.075em);
    \ifx\xvec@arg\xvec@d%
      \fill(\xvec@width*.45,.5ex) circle (.5pt);%
    \else\ifx\xvec@arg\xvec@dd%
      \fill(\xvec@width*.30,.5ex) circle (.5pt);%
      \fill(\xvec@width*.65,.5ex) circle (.5pt);%
    \fi\fi%
    \end{tikzpicture}%
  }}}%
  #2%
}
\makeatother

\newlist{todolist}{itemize}{2}
\setlist[todolist]{label=$\square$}
\usepackage{pifont}
\def\Vecq{\mathbf{q}}

\def\Vecu{\mathbf{u}}
\def\Vecv{\mathbf{v}}

\def\VecT{\mathbf{T}}
\def\VecU{\mathbf{U}}
\def\VecV{\mathbf{V}}

\begin{document}

\title{Threading light through dynamic complex media}

\author{Chaitanya~K.~Mididoddi}
\email{c.mididoddi@exeter.ac.uk}
\affiliation{Physics and Astronomy, University of Exeter, Exeter, EX4 4QL. UK.}
\author{Christina Sharp}
\affiliation{Physics and Astronomy, University of Exeter, Exeter, EX4 4QL. UK.}
\author{Philipp del Hougne}
\affiliation{Univ.\ Rennes, CNRS, IETR -- UMR 6164, F-35000 Rennes, France.}
\author{Simon~A.~R.~Horsley}
\affiliation{Physics and Astronomy, University of Exeter, Exeter, EX4 4QL. UK.}
\author{David~B.~Phillips}
\email{d.phillips@exeter.ac.uk}
\affiliation{Physics and Astronomy, University of Exeter, Exeter, EX4 4QL. UK.}

\begin{abstract}
The scattering of light impacts sensing and communication technologies throughout the electromagnetic spectrum. Overcoming the effects of {\it time-varying} scattering media is particularly challenging. In this article we introduce a new way to control the propagation of light through dynamic complex media. Our strategy is based on the observation that many dynamic scattering systems exhibit a range of decorrelation times -- meaning that over a given timescale, some parts of the medium may essentially remain static. We experimentally demonstrate a suite of new techniques to identify and guide light through these networks of static channels -- threading optical fields around multiple dynamic pockets hidden at unknown locations inside opaque media. We first show how a single stable light field propagating through a partially dynamic medium can be found by optimising the wavefront of the incident field. Next, we demonstrate how this procedure can be accelerated by 2 orders of magnitude using a physically realised form of adjoint gradient descent optimisation. Finally, we describe how the search for stable light modes can be posed as an eigenvalue problem: we introduce a new optical matrix operator, the {\it time-averaged transmission matrix}, and show how it reveals a basis of {\it fluctuation-eigenchannels} that can be used for stable beam shaping through time-varying media. These methods rely only on external camera measurements recording scattered light, require no prior knowledge about the medium, and are independent of the rate at which dynamic regions move. Our work has potential future applications to a wide variety of technologies reliant on general wave phenomena subject to dynamic conditions, from optics to acoustics.
\end{abstract}

\maketitle

\noindent{\bf Introduction}\\
Optical scattering randomly redirects the flow of light. It is a ubiquitous phenomenon that has wide-ranging effects. Since imaging relies on light travelling in straight lines from a scene to a camera, scattering prevents image formation through fog, and precludes high-resolution microscopy inside biological tissue~\cite{bertolotti2022imaging,gigan2022roadmap}. Scattering also impairs optical communications through air and water, and disrupts the transmission of microwave and radio signals~\cite{saigre2022self}. Overcoming the adverse effects of light scattering is an extremely challenging problem~\cite{carminati2021principles}. Nonetheless, due to its prominence, substantial progress has been made over the last decades~\cite{rotter2017light}.

When light propagates through a strongly scattering medium (also known as a `complex' medium~\cite{bertolotti2022imaging}), the wavefront of the incident optical field is distorted, corrupting the spatial information it carries. Elastic scattering from a {\it static} medium is deterministic, meaning that the precise way in which light has been perturbed can be characterised and subsequently corrected. By sending a series of probe measurements through the medium, a digital model of its effect on light can be created: represented by a linear matrix operator known as a transmission matrix (TM)~\cite{popoff2010measuring}. Once measured, the linearity of Maxwell's equations means that the TM describes how any linear combination of the probe fields will be transformed. The TM reveals how to best undo the distortion imparted to a scattered field emerging from a complex medium, and the time-reverse: how to pre-distort an input optical field so that it evolves into a user-defined state at the output -- a technique known as wavefront shaping~\cite{vellekoop2007focusing}.

Using modern high-resolution spatial light modulators (SLMs), it is possible to precisely measure and control the relative intensity, phase and polarization of thousands of independent optical spatial modes as they undergo many scattering events inside a highly turbid medium~\cite{mosk2012controlling}. Thus, wavefront shaping, and the closely related technique of optical phase conjugation~\cite{yaqoob2008optical}, have been used to image up to a depth of several hundred microns into fixed biological tissue~\cite{papadopoulos2017scattering}. TM-based approaches have also inspired new forms of ultra-thin micro-endoscopy through rigidly-held strands of multimode optical fibre (MMF)~\cite{turtaev2018high}.

Despite these successes, control of light through {\it time-varying} complex media remains a largely open problem~\cite{gigan2022roadmap}. Evidently, wavefront shaping can only be successfully applied if the medium in question remains predominantly stationary for the time taken to make probe measurements and apply a wavefront correction. Yet many application scenarios feature complex media that rapidly fluctuate on a timescale of milliseconds or faster -- rendering wavefront shaping approaches exceedingly difficult~\cite{jang2015relation}. Overcoming this challenge offers a stepping stone to a potent array of new technologies, including the ability to look directly inside living biological tissue, to see through fog, and to increase the data-rate of optical communications through the turbulent atmosphere and flexible fibre-optics.

So far, the main strategies to control light through moving complex media have focused on achieving the task of wavefront shaping as quickly as possible~\cite{liu2017focusing,blochet2017focusing,yang2020fighting,may2021fast,luo2022high}. In the optical regime, beam shaping at kiloHertz rates can be implemented with digital micro-mirror devices (DMDs)~\cite{conkey2012high,wang2015focusing,turtaev2017comparison}. The need for yet higher switching rates has spawned the development of ultra-fast SLMs capable of wavefront shaping at hundreds of kiloHertz~\cite{Tzang2019,feldkhun2019focusing} while megaHertz to gigaHertz modulation-rate SLMs hold future promise~\cite{peng2019design,panuski2022full}. Spectral multiplexing enables many probe measurements to be made simultaneously, speeding up the data gathering part of the wavefront shaping process~\cite{feldkhun2019focusing,wei2020real}. In addition, the number of probe measurements needed to reconstruct a usable TM can be reduced by exploiting prior knowledge about the medium itself -- such as correlations between elements of the TM (known as memory effects), predictions about how the power is distributed over the TM elements, or a recent but slightly degraded TM measurement~\cite{bertolotti2012non,judkewitz2015translation,horstmeyer2015guidestar,osnabrugge2017generalized,badon2020distortion,li2021memory,li2021compressively,valzania2022online}. Fast optical focusing inside biological tissue can be achieved with optical phase conjugation guided by ultrasonic guide-stars -- relying on the lower levels of scattering experienced by ultrasound~\cite{wang2012deep,judkewitz2013speckle,lai2015photoacoustically,katz2019controlling}. A variety of other methods relying on correlations between the object of interest and externally measurable signals offer alternative routes to image through moving complex media~\cite{paniagua2019blind,ruan2020fluorescence}.

Here we introduce a new way to control the propagation of light through dynamic scattering media. Our approach is complementary to existing techniques. We begin by classifying complex media into three categories, based on the level and type of motion exhibited over the timescale required for wavefront shaping, denoted by $\tau_{\text{ws}}$. 
Class 1 represents static complex media that remain completely fixed over time $\tau_{\text{ws}}$. Established TM-based methods can be applied to deterministically control scattered light in this case. Class 2 represents moving complex media, which undergo substantial motion everywhere over time $\tau_{\text{ws}}$. This class of media eludes current wavefront shaping approaches. However, there is an opportunity to make progress by considering a third class -- representing an edge-case between classes 1 and 2. Class 3 comprises {\it partially} moving scattering media, which, over the timescale $\tau_{\text{ws}}$, exhibit localised pockets with time-varying properties embedded within a static medium. Any dynamic complex medium possessing a range of decorrelation rates has the potential to be classified in this way. For example, this situation describes: tissue in which small capillaries conducting blood flow represent faster moving regions surrounded by a matrix of more slowly changing scattering material; pockets of turbulent air above hot chimneys within calmer air over a city skyline; and the movement of people modifying the scattering of microwaves only at floor level throughout a building.

In this article we focus on how to identify light fields that predominantly stay within the static regions of such partially moving complex media (i.e.\ class 3 media). We experimentally demonstrate three new techniques that excite largely stable modes within these environments. We show how these optimised modes scatter almost entirely around all moving pockets. These methods do not rely on prior knowledge of the location of dynamic regions and only require measurements external to the medium. These measurements can be made on the same timescale or more slowly than the medium is fluctuating -- crucial for the practical application of these techniques. Our work expands the toolkit of methods to overcome dynamic scattering, pointing to a range of future applications in the fields of imaging, optical communications, and beyond.\\

\begin{figure*}[t]
   \includegraphics[width=1\textwidth]{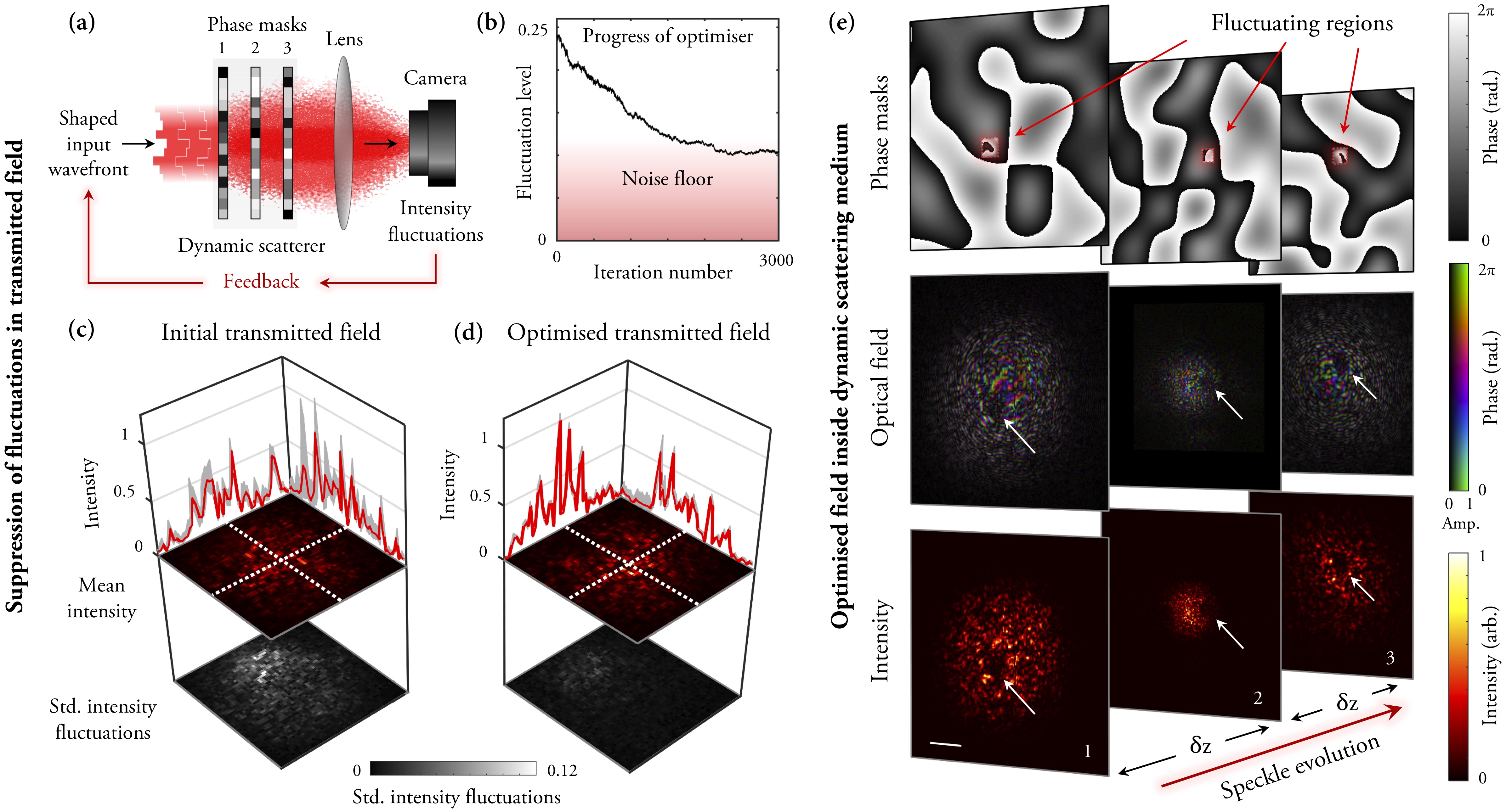}
   \caption{{\bf Unguided optimisation}. (a) Schematic of experimental set-up. An input wavefront is iteratively modified to reduce the intensity fluctuations in transmitted light. (b) A plot of fluctuation level as a function of iteration number throughout the optimisation procedure. Convergence is reached after several thousand iterations: the fluctuation level does not fall to zero, but plateaus when the residual fluctuations fall below the experimental noise floor, indicated (approximately) in pink. (c) Fluctuations in the output field for a randomly chosen input field used as the starting point of the optimisation. Upper heat maps show the mean intensity of transmitted light at the output plane, and lower heat maps show the fluctuation level around the mean, represented as a standard deviation around the mean. The line-plots show line-profiles through the output field along the lines marked with white hatched lines, with mean intensity (red line) and fluctuations about the mean (gray shading). (d) Equivalent plot to (c) but now showing the optimised transmitted field. We see the fluctuations have been strongly suppressed in (d) compared to (c). (e) Measured shape of the optimised field inside the dynamic scattering sample. The top row shows the 3 phase planes that form the scattering system, with a fluctuating region on each plane highlighted by a red box. The middle and bottom rows show the optical field (middle row) and intensity pattern (absolute square of the field -- bottom row) incident on each plane. We see that the optimised field arriving at each plane has a low intensity region corresponding to the location of the fluctuating region -- highlighted by white arrows -- thus `avoids' these regions.}
   \label{Fig:unguided}
\end{figure*}

\noindent{\bf Results}\\
When a light field $\Vecu$ is incident on a time-varying medium, the time-dependent transmitted field is given by
\begin{equation}
    \Vecv(t) = \VecT(t)\boldsymbol\Vecu,
\end{equation}
where $\VecT(t)$ is the time-dependent transmission matrix of the medium, and here $\Vecu$ and $\Vecv$ are represented as column vectors. Our aim is to find an input $\Vecu$ that scatters around dynamic regions within the medium, thus minimising the fluctuations in the output field $\Vecv(t)$.

To experimentally investigate this new form of light control, we emulate a three-dimensional time-varying scattering medium using a cascade of three computer controlled diffractive optical elements, each separated by a free-space distance of $\delta z$.  Cascades of phase planes can emulate atmospheric turbulence~\cite{kolmogorov1941local,cox2020structured,bachmann2022highly} and have also been shown to mimic the complicated optical scrambling effects of a multiple scattering sample~\cite{boucher2021full,kupianskyi2022high}. In practice this set-up is implemented using multiple reflections from a single liquid crystal SLM, allowing the phase profiles to be arbitrarily digitally reconfigured. We choose this test-bed as it is a versatile way to control the degree of scattering, and the number and location of dynamic regions for proof-of-principle experiments.

As shown in Fig.~\ref{Fig:unguided}(e), top row, we display a static random phase pattern on each phase screen, spatially distorting optical signals flowing through the optical system. On each plane we also define an area within which the phase profile is programmed to randomly fluctuate in time -- these patches represent the `pockets' of dynamic material embedded inside the scattering sample. A second SLM is used to shape the light incident onto the dynamic medium, and a camera records the level of intensity fluctuations in transmitted light.\\

\noindent{\bf Unguided optimisation:} We first pursue a straight-forward optimisation method: iterative modification of input field $\Vecu$ to suppress intensity fluctuations at the output. Figure~\ref{Fig:unguided}(a) shows a schematic of this approach. Supplementary Information (SI) \S1 shows a full diagram of the optical set-up. The optimisation commences by transmitting an initial trial field $\Vecu_0$ through the sample, and recording the intensity fluctuations on the camera. We sample 20 realisations of the fluctuating speckle pattern, and the level of fluctuations over these frames is quantified by the objective function $F = \Bar{\sigma}_I/\Bar{I}$, where $\Bar{\sigma}_I$ denotes the standard deviation of the fluctuating intensity, averaged over all illuminated camera pixels, and $\Bar{I}$ is the average transmitted intensity. This choice of objective function ensures that fluctuations are normalised with respect to transmitted power.

The input SLM used to generate the incident fields is subdivided into 1200 super-pixels. The phase delays imparted by these super-pixels represents the independent degrees-of-freedom we aim to optimise. We begin by setting each super-pixel to a random phase value, creating incident field $\Vecu_0$, and measure the level of output fluctuations. Next, two new test fields are sequentially transmitted through the sample. These are generated by randomly selecting half of the input SLM super-pixels used to create $\Vecu_0$, and adding/subtracting a small constant phase offset $\delta\theta$ from these pixels, yielding inputs $\Vecu_{\pm\delta\theta}$. We measure the corresponding level of output fluctuations for these two new trial inputs, and if either exhibit lower fluctuations than $\Vecu_0$, the optimised input field is updated accordingly. This process is repeated until the output fluctuation level no longer improves.

This algorithm relies on accurately capturing the output fluctuations on each iteration. However, even in the absence of any other sources of noise, there is an uncertainty in the measurement of $\bar{\sigma}_I$ and $\bar{I}$ due to the finite number of realisations of the dynamic medium sampled. To enhance the algorithm's robustness to this source of noise, on each new iteration we re-test the optimum input field from the last iteration and compare this to the new trial fields -- doing so increases the optimisation time, but crucially prevents a single measurement with an erroneously low value of $F$ from blocking the optimiser from taking steps in subsequent iterations. Figure~\ref{Fig:unguided}(b) shows a typical optimisation curve throughout our experiment. The noise floor is governed by the uncertainty in real fluctuations detailed above, along with small variations in the intensity of the laser source, camera noise and uncontrolled fluctuations in light reflecting from the liquid crystal SLM as it is updated, which all add to the apparent level of measured fluctuations.

\begin{figure*}[t]
   \includegraphics[width=1\textwidth]{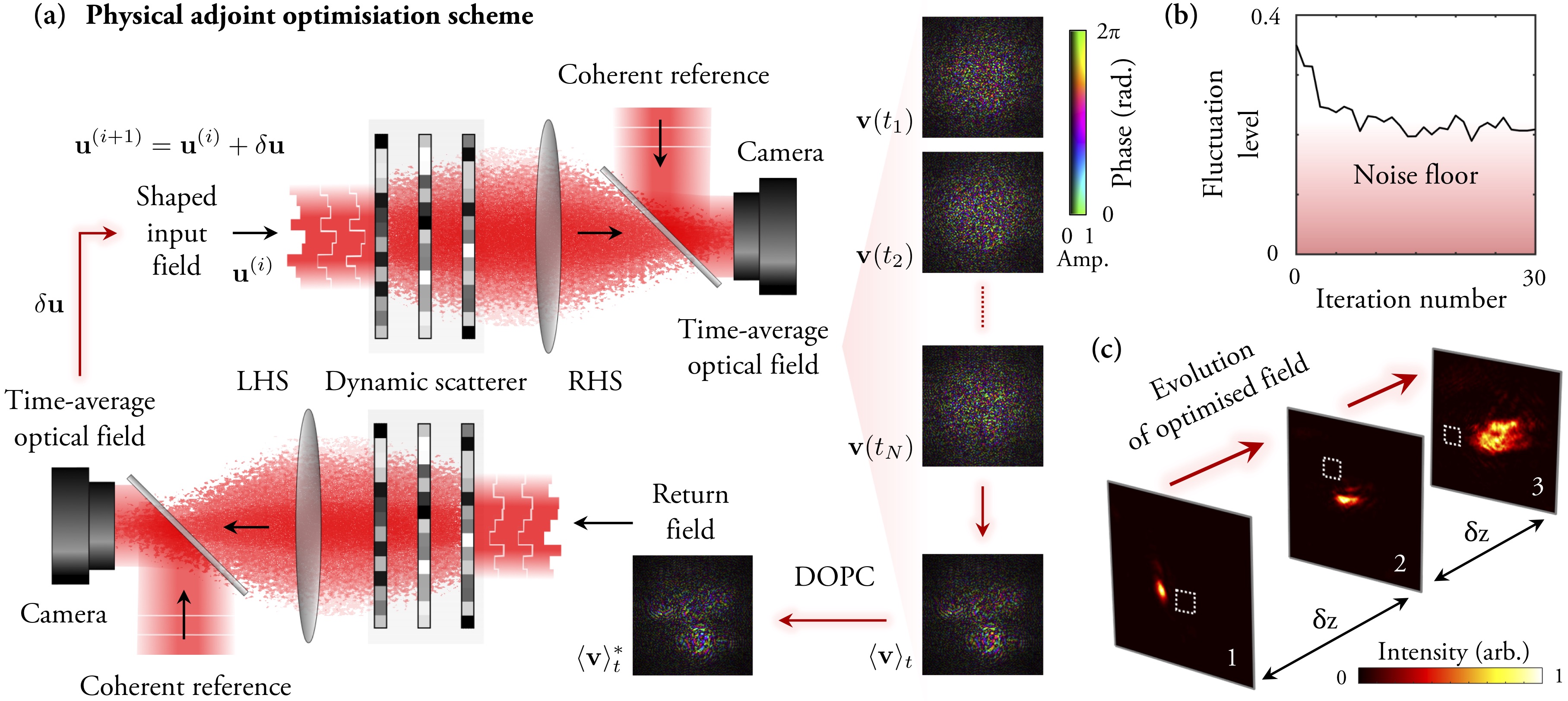}
   \caption{{\bf Physical adjoint optimisation}. (a) Schematic of experimental set-up. On iteration $i$ an input field $\Vecu^{(i)}$ is transmitted through the dynamic medium from the left-hand-side (LHS). The output field is time-averaged on the right-hand-side (RHS) -- the schematic shows output fields recorded at individual times $\Vecv(t_1)$, $\Vecv(t_2)\dotsb\Vecv(t_N)$ (where $N$ is the total number of recorded output fields). These are averaged to yield $\left<\Vecv\right>_t$. Digital optical phase conjugation (DOPC) is carried out to transmit the phase conjugate of $\left<\Vecv\right>_t$ back through the medium. The resulting field emerging on the LHS is then time-averaged, and used to calculate $\delta\Vecu$, such that the input of the next iteration ($i+1$) is given by $\Vecu^{(i+1)}=\Vecu^{(i)}+\delta\Vecu$. (b) A plot of fluctuation level as a function of iteration number throughout the optimisation procedure. In this scheme, convergence is reached after $\sim15$ iterations. (c) The experimentally recorded intensity of the optimised field arriving at the three phase planes. The maximum intensity at each plane is normalised to 1. The white squares indicated the location of the moving region on each plane. We see that, once again, the optimised field avoids these moving regions of the sample.}
   \label{Fig:DOPC}
\end{figure*}
Figures~\ref{Fig:unguided}(c) and~\ref{Fig:unguided}(d) show examples of the output fluctuations of an initial trial field (c) and an optimised field (d) using this approach. See also Supplementary Movie 1. We see that fluctuations of the output field are heavily suppressed after optimisation. As we have full control over the test scattering medium, we are able to digitally `peel back' the outer scattering layers to look inside and directly observe the evolution of the optimised field as it propagates through the cascade of phase planes. Experimentally this is achieved by switching-off the aberrating effect of the second and third phase planes, and imaging the optimised field that is incident on plane 2. We recover the phase of this optical field using digital holography, and reconstruct the fields at planes 1 and 3 by numerically propagating the field at plane 2 (see SI \S2). We see the optimised field scatters through the medium to form a speckle pattern that evolves to exhibit near-zero intensity at the locations of the fluctuating regions on each plane (Fig.~\ref{Fig:unguided}(e), bottom row) -- thus avoiding these dynamically changing regions and minimising fluctuations in the transmitted field.

This is an encouraging result, however this form of undirected optimisation is a relatively slow process -- in this case requiring several thousand iterations to converge (see Fig.~\ref{Fig:unguided}(b)). Therefore, we next ask: is there a way to find optimised fields more rapidly?\\

\noindent{\bf Physical adjoint optimisation:} In our first strategy, on each iteration we directly measure how one randomly chosen spatial component of the input field should be adjusted to reduce the fluctuations in the output field. We now describe a more sophisticated technique to simultaneously obtain how all spatial components composing the input field should be adjusted in parallel. This strategy converges to an optimised input beam in far fewer iterations than unguided optimisation.

Our approach can be understood as gradient descent optimisation using fast adjoint methods. Adjoint optimisation refers to the efficient computation of the gradient of a function for use in numerical optimisation. Here, we lack sufficient information to numerically perform this adjoint operation, but instead we show how it is possible to {\it physically} realise it by passing light in both directions through the dynamic scattering medium.

SI \S3 gives a detailed derivation of this method. In summary, to suppress output fluctuations we aim to maximise the correlation (i.e.\ overlap integral) between all output fields over time, given by the real positive scalar objective function
\begin{equation}
    F = \left|\sum^T_{t=1}\sum^T_{t'=1}\left[\Vecv^\dagger(t)\cdot\Vecv(t')\right]\right|^2.
\end{equation}
To increase $F$, at each iteration we incrementally adjust the complex field of all elements of the input field $\Vecu$, so that the input field at iteration $i+1$ is given by $\Vecu^{(i+1)} = \Vecu^{(i)}+\delta\Vecu$, where $\Vecu^{(i)}$ is the input field of iteration $i$, and column vector $\delta\Vecu = \delta A{\textrm e}^{i\boldsymbol{\theta}}$.
Here $\delta A$ is the optimisation step size: a small real positive constant, and we find (see SI \S3) that column vector $\boldsymbol{\theta}$ is given by
\begin{equation}\label{Eq:conjAv}
    \boldsymbol{\theta} = -\arg\left(\VecT^T\cdot\langle\Vecv^*\rangle_t\right),
\end{equation}
where $\langle\Vecv^*\rangle_t$ is the phase conjugate of the time-averaged output field.

Our adjoint optimisation scheme is shown schematically in Fig.~\ref{Fig:DOPC}(a). Iteration $i$ commences by illuminating the dynamic scattering medium from the left-hand-side (LHS) with trial field $\Vecu^{(i)}$, and time-averaging the transmitted optical field on the right-hand-side (RHS), yielding $\langle\Vecv\rangle_t$. Equation~\ref{Eq:conjAv} specifies that $\langle\Vecv\rangle_t$ should be phase conjugated, and transmitted in the reverse direction through the dynamic media, from the RHS back to the LHS. Measuring the phase of the resulting field on the LHS yields information about how all spatial components of the input field should be modified to improve $F$, enabling calculation of the next input $\Vecu^{(i+1)}$.

\begin{figure*}[t]
   \includegraphics[width=1\textwidth]{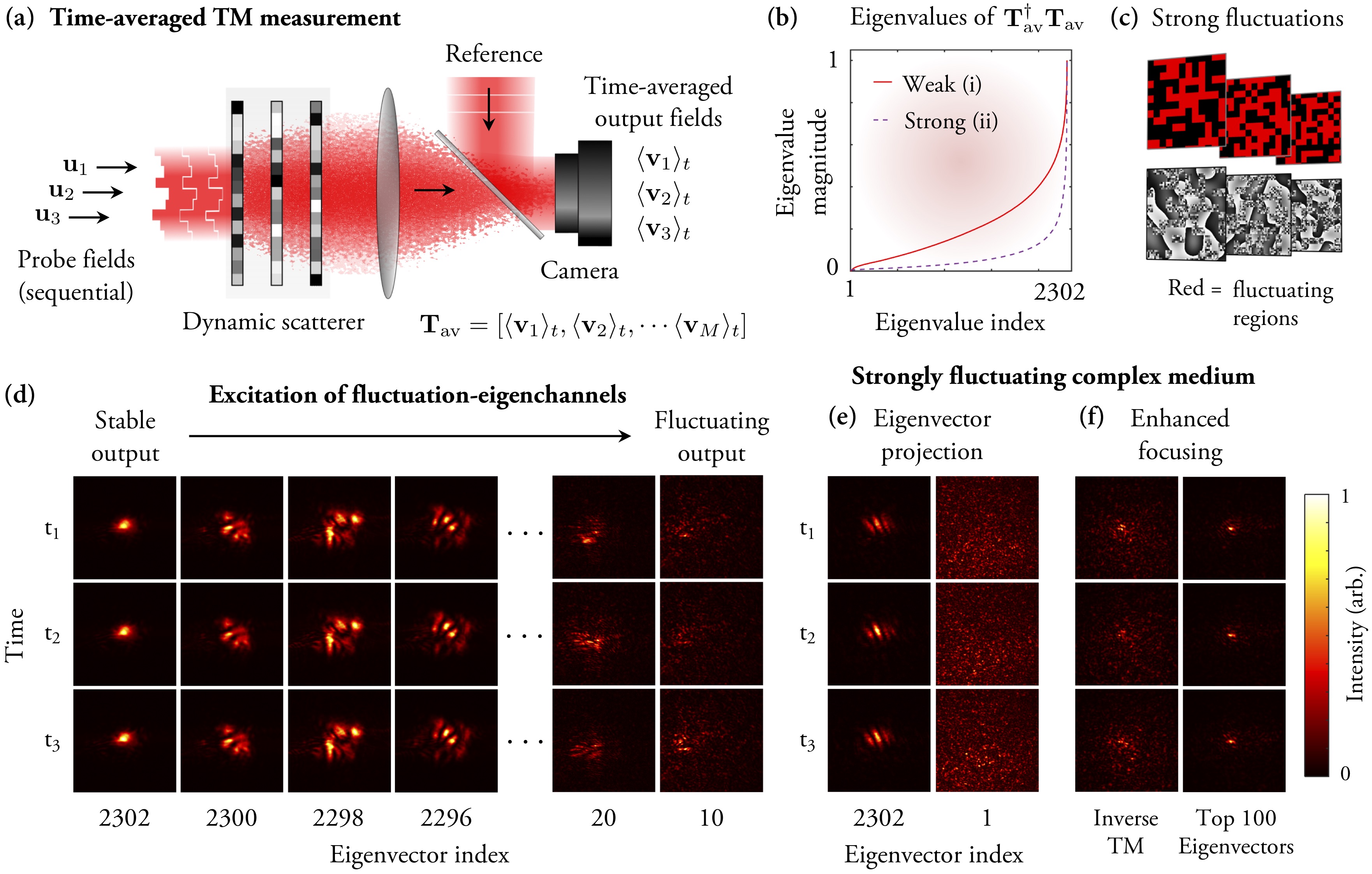}
   \caption{{\bf Time-averaged transmission matrix}. (a) Schematic of experimental set-up. A sequence of orthogonal probe fields are individually transmitted through the medium, e.g.\ $\Vecu_1$, $\Vecu_2$, $\Vecu_3$. For each input, the corresponding time-averaged output field is recorded, e.g.\ $\left<\Vecv_1\right>$, $\left<\Vecv_2\right>$, $\left<\Vecv_3\right>$, and arranged column-by-column to build the time-averaged TM $\VecT_{\text{av}}$. (b) The magnitudes of the eigenvalues of $\VecT_{\text{av}}^\dagger\VecT_{\text{av}}$, for a weakly (i) and strongly (ii) fluctuating dynamic medium. Both are arranged in ascending order and normalised to a maximum value of 1. The weakly fluctuating medium is the same as used in the earlier experiments. An example of the strongly fluctuating medium is shown in (c), with moving regions highlighted in red. (d) Excitation of selected fluctuation-eigenchannels in the weakly fluctuating medium. Each column shows the output when the medium is illuminated with different eigenvectors. Each row shows the output at a different time -- i.e.\ for 3 different configurations of the dynamic regions of the medium. We see the high index eigenvectors are stable with respect to these movements, while the low eigenvectors are not. (e) Eigenvector projection through a strongly fluctuating medium. (f) Enhanced focusing through strongly fluctuating scattering media using the time-averaged TM. Left column: an attempt to make a focus using the conventional inverse TM, which is measured while the medium fluctuates. We see a poor contrast focus which fluctuates strongly as the medium reconfigures. Right column: An output focus created through the same medium, with the input field generated using the top 100 most stable eigenvectors of $\VecT_{\text{av}}^\dagger\VecT_{\text{av}}$. Here we see that the contrast and stability of the output focus is significantly improved.}
   \label{Fig:TimeAv}
\end{figure*}

Experimentally, this adjoint field optimisation strategy requires a relatively complicated optical setup: two digital optical phase conjugation (DOPC) systems -- which enable time-reversal of optical fields -- are arranged back-to-back on either side of the dynamic sample. We use single-shot off-axis digital holography to measure the output fields on each side. The DOPC systems require very precise alignment, so we implemented a calibration method that we recently described in ref.~\cite{mididoddi2020high}. Our set-up enables spatial shaping of both the intensity and phase profile of time-reversed field travelling in both directions. We test this approach to guide light through a similar sample dynamic medium to before (see Fig.~\ref{Fig:unguided}(e), top row) and average over $N=5$ realisations of the medium in each direction. SI \S4 shows a schematic of the full optical set-up used in this experiment.

Figure~\ref{Fig:DOPC}(b) shows a typical convergence curve throughout the optimisation process. After only $\sim15$ iterations, the input field converges to a solution with output fluctuations suppressed to a similar level as unguided optimisation -- crucially now achieved in over 2 orders of magnitude fewer iterations. Supplementary Movie 2 shows the output fluctuations before and after optimisation. Once again looking inside the dynamic sample, we see that we have found a more localised optical field that passes almost entirely through the static parts of each phase plane and avoids the moving regions, as shown in Fig.~\ref{Fig:DOPC}(c).\\

\noindent{\bf The fluctuation-eigenchannels of the time-averaged TM}: So far we have focused on strategies to find a single optimised input field as quickly as possible. We now consider how a set of input modes may be determined, that all navigate around moving regions of a dynamic medium. Knowledge of such a sub-basis would enable a stable shaped output field -- such as a focused spot -- to be formed from a suitable linear combination of these time-independent fields at the output plane. This opens up the prospect of imaging through partially dynamic scattering media.

One possibility is to conduct a series of adjoint optimisations, each seeded from a different initial field. This would lead to a set of stable output fields that can be stored as the column vectors of matrix $\VecV$, and used to generate a target output field $\Vecv_{\text{trg}}$ by injecting into the medium the field $\Vecu = \VecV^{-1}\Vecv_{\text{trg}}$. However, this is not an efficient search strategy, since there is no way to guarantee the linear independence of the set of optimised fields -- meaning very similar fields may be inadvertently found.

To overcome this problem, we now devise a new method capable of finding the full set of orthogonal fields that navigate around moving regions, for a given input basis. We make use of the information stored in the {\it time-averaged transmission matrix} of a fluctuating optical system: $\VecT_{\text{av}}$. To measure $\VecT_{\text{av}}$, a set of $M$ probe fields are sequentially transmitted through the dynamic sample, and the time-averaged output field is calculated in each case, forming the columns of $\VecT_{\text{av}}$. Figure~\ref{Fig:TimeAv}(a) shows a schematic of this approach. We illuminate the sample with $M = 2304$ probe fields, and average the output field over $N = 10$ uncorrelated realisations of the scattering medium for each input mode. Experimentally this procedure is simpler than physical adjoint optimisation -- although the main challenge is that the reference beam required for holographic field measurement must be phase-drift-stabilised for the entire measurement of $\VecT_{\text{av}}$. SI \S5 describes the full optical setup for this experiment.

We aim to discover fields that deliver high levels of time-averaged energy to the output plane. Finding these fields can be represented as an eigenvalue problem by noting that the total intensity $P$ arriving at the output in field $\Vecv$ can be expressed as
\begin{equation}
    P = \Vecv^\dagger\Vecv = \Vecu^\dagger\VecT_{\text{av}}^\dagger\VecT_{\text{av}}\Vecu.
\end{equation}
Therefore, the eigenvectors of matrix $\VecT_{\text{av}}^\dagger\VecT_{\text{av}}$ with the largest absolute eigenvalues represent input fields that deliver the highest time-averaged intensity to the output plane. Assuming the internal fluctuations of the medium are large enough to randomise the phase of scattered light, then the fluctuating parts of the output fields will average to near-zero. When forward scattering dominates, eigenvectors with high absolute eigenvalues also correspond to input fields that interact least with the time-varying regions inside the medium (i.e.\ these fields are the least `averaged away'). We term this basis of eigenvectors the {\it fluctuation-eigenchannels} of the dynamic medium.

Figure~\ref{Fig:TimeAv}(b) shows the distribution of absolute eigenvalues of the matrix $\VecT_{\text{av}}^\dagger\VecT_{\text{av}}$, arranged in ascending order. Here we compare the eigenvalue distribution resulting from time-averaged TMs measured on two independent dynamic samples with a different numbers of moving regions: (i) has a single dynamic patch on each plane similar to that shown in Fig.~\ref{Fig:unguided}; (ii) has randomly placed fluctuating patches covering approximately half of the area of each plane -- an example is shown in Fig.~\ref{Fig:TimeAv}(c). In Fig.~\ref{Fig:TimeAv}(b) we see that the magnitudes of the eigenvalues decrease more steeply from the maximum value in this second case, indicating that the spectrum of eigenvectors deliver less time-averaged energy to the output -- i.e.\ there are fewer stable channels available through a sample with more extensive moving regions, as would be expected.

We first demonstrate excitation of the fluctuation-eigenchannels of the more weakly fluctuating sample medium (i). Figure~\ref{Fig:TimeAv}(d) shows examples of output speckle patterns when a selection of fluctuation eigenchannels are excited, with some of the highest and lowest absolute eigenvalues. Each row shows the output field for a new configuration of the dynamic medium (recorded at distinct times $\text{t}_1$, $\text{t}_2$, $\text{t}_3$). The transmitted fields corresponding to high index fluctuation-eigenchannels remain stable (i.e.\ largely unchanging), indicating that the light propagating through the medium in these cases is avoiding dynamic regions. Conversely, the transmitted fields corresponding to low index eigenchannels vary with time at the output -- as these modes interact strongly with the moving parts of the dynamic sample. Supplementary Movie 3 shows examples of the stability of output light transmitted through a range of different fluctuation-eigenchannels.

We now investigate light shaping capabilities through the more challenging strongly fluctuating medium (ii). Figure~\ref{Fig:TimeAv}(e) shows the stability of transmitted fields when exciting fluctuation-eigenchannels with the highest (left column) and lowest (right column) absolute eigenvalues. In this case, even light propagating through the most stable eigenchannel exhibits non-negligible output fluctuations over time, indicating that we have not found any fields that thread perfectly around all moving parts of the sample. Despite this, we find that a significant improvement in focusing at the output is possible using the information stored in the time-averaged TM. Figure~\ref{Fig:TimeAv}(f) shows a focus created using a conventional TM approach, where the medium freely fluctuates throughout TM measurement (left column) compared to using a sub-basis formed from the top 100 most stable fluctuation-eigenchannels (right column) -- see SI \S6 for details, and Supplementary Movie 4. We see that both the contrast and stability of the focus is strongly enhanced using our new approach.\\

\noindent{\bf Discussion and conclusions}\\
In summary, we have identified a broad new class of {\it partially} dynamic scattering media that is amenable to deterministic light control techniques. We have demonstrated three new ways to thread stable light fields through such media, that rely on the movement of the medium itself to accomplish:

The first technique, unguided optimisation, is a straight-forward but relatively slow approach, most suited to the case where the network of static channels throughout the medium remains fixed. Here our optimisation strategy is analogous to the first methods used to shape light through static scattering media~\cite{vellekoop2007focusing}, and as such may be improved using more advanced algorithms~\cite{vellekoop2008phase,conkey2012genetic}. This technique is also highly flexible: the form of the objective function can be arbitrarily chosen. For example, intensity shaping terms could also be included, to simultaneously reduce fluctuations and shape the output.

The second approach, physical adjoint optimisation, enables stable light fields to be very rapidly found by passing light backwards and forwards through the medium. We {\it physically compute} the gradient of the objective function with respect to the optimisation variables (i.e.\ the field emanating from each super-pixel on the SLM). If implemented with fast beam shaping, this technique is well-suited to the case where a particular configuration of static channels only persist for a relatively short time. Our adjoint strategy is reminiscent of iterative time-reversal~\cite{prada1995iterative}, and recently proposed in-situ methods to train photonic neural networks~\cite{hughes2018training}. Indeed our work may be considered one of the first real-world implementations of a photonic adjoint optimisation routine -- a challenging yet powerful method to realise experimentally~\cite{zhou2020situ}. We note that, for our application, the form of objective function is more restricted than unguided optimisation. For example, we found that some choices of objective function require deterministic control over the motion of the dynamic parts of the scattering medium which is evidently not possible in most cases.

Our final strategy relies on measurement of the time-averaged TM to calculate the fluctuation-eigenchannels of the dynamic medium. These channels are excited by an orthogonal set of input eigenfields, ordered in terms of how much time-averaged power they deliver to the output plane -- thus revealing internal fields that minimally interact with the time-varying parts of the medium. This new concept is related to several previously introduced matrix operators connected to physical quantities of interest in scattering media
~\cite{popoff2010measuring,ambichl2017focusing,durand2019optimizing,bender2022depth}. As fluctuations in the medium go to zero, the time-averaged TM becomes equivalent to the conventional TM, and the fluctuation-eigenchannels tend to the transmission-eigenchannels of a static scattering medium~\cite{kim2012maximal}. The `deposition matrix'~\cite{bender2022depth} and the `generalised Wigner-Smith operator'~\cite{ambichl2017focusing,matthes2021learning} are both also capable of revealing light fields that circumnavigate predetermined regions within a complex medium. However, only the time-averaged TM does so without requiring access to internal fields within the medium~\cite{bender2022depth} or the measurement of an entire TM while the medium is held static~\cite{ambichl2017focusing}. Here we have demonstrated that the time-averaged TM can enhance focusing through partially dynamic scattering media. We also expect an improvement in more elaborate beam shaping, such as point-spread function engineering~\cite{boniface2017transmission} and arbitrary pattern projection~\cite{li2021compressively}.

We note that previous studies have used localised internal motion within scattering media as a guide-star -- enabling focusing directly onto these moving regions~\cite{zhou2014focusing,Ruan:17} -- the opposite of what we have set out to achieve in our study. Recent work also investigated the performance of wavefront optimisation occurring on the same timescale as the medium decorrelates -- with evidence to suggest that the resulting focus was dominated by the most stable modes propagating through the medium~\cite{blochet2019enhanced}.

In this study, our experiments have emulated mainly anisotropic forward-scattering media, as would be found transmitting light through the atmosphere, through multimode fibres, or through thin layers of biological tissue. In the future it will be interesting to study how these techniques perform, or indeed need to be adapted, as the level of multiple scattering increases to the onset of the diffusive~\cite{cao2022shaping} or chaotic regimes~\cite{del2020robust}. While we expect the strategies outlined here to apply in these domains, strongly scattering environments also pose additional challenges, since there are competing requirements: optical fields must both circumnavigate moving regions and also penetrate deeply enough into the medium to transmit significant power to the other side. We expect a smaller number of internal fields will satisfy both of these constraints, since optimised fields will be formed from a reduced basis of modes -- dominated by high transmission-eigenchannels that are weighted to destructively interfere on all moving regions~\cite{bender2022depth}. A further avenue of exploration would be to investigate optimal focusing {\it inside} partially dynamic media, while simultaneously guiding light around moving pockets.

Finally, it is interesting to note that the problem we have addressed in our work, from an optical perspective, is closely related to the concept of multi-path fading experienced by radio frequency wireless communication channels. In this latter case the interaction of transmitted signals with moving media in their path is known as {\it mode-stirring}, and the {\it Rician K-factor} quantifies the ratio of `unstirred' to `stirred' paths transmitted through a dynamic environment~\cite{hosli2004capacity,taricco2006cth09,simon2008digital}. Circumnavigation of localised dynamic regions of space may potentially be applied at radio and microwave frequencies, either in the spectral domain, or in the spatial domain in conjunction with beam-forming systems.

The concepts that we have introduced here apply generally to wave phenomena, and have relevance to a diverse range of applications. Possibilities include imaging deep into living biological tissue~\cite{booth2014adaptive}, transmission of space-division multiplexed optical communications through turbulent air~\cite{zhu2002free} and underwater~\cite{kaushal2016underwater}, propagation noise reduction in acoustic beam forming~\cite{chiariotti2019acoustic} and emerging smart microwave and radio environments~\cite{renzo2019smart}. Our work adds to the toolbox of methods to counteract the adverse effects of dynamic scattering media.\\

\noindent{\bf Author contributions}\\
DBP conceived the idea for the project and developed it with all other authors. CKM performed all experiments and data analysis, with support from DBP. SARH derived the physical adjoint optimisation method. DBP and PdH conceived the time-averaged TM method. CS and CKM performed supporting simulations, with guidance from DBP and SARH. DBP wrote the manuscript with editorial input from all other authors.\\

\noindent{\bf Acknowledgements}\\
SARH acknowledges the Royal Society and TATA for financial support through grant URF\textbackslash R\textbackslash 211033. DBP acknowledges the financial support of the Royal Academy of Engineering and the European Research Council (Grant no. 804626: `Rendering the opaque transparent'). We thank Jacopo Bertolotti for useful discussions.\\

\noindent{\bf Competing interest}\\
The authors declare no competing interests.\\

\noindent{\bf Data availability}\\
Additional information is available in Supplementary Information. The raw data that support the results in this article will be made available at the University of Exeter repository (https://ore.exeter.ac.uk/repository/)
upon publication. Alternatively, these data are available from the corresponding author upon reasonable request.

\bibliography{dynamicTMRefs}

\onecolumngrid
\newpage

\hspace{6mm}{\Large Threading light through dynamic complex media: supplementary information}\\

\begin{figure*}[b]
  \includegraphics[width=0.7\textwidth]{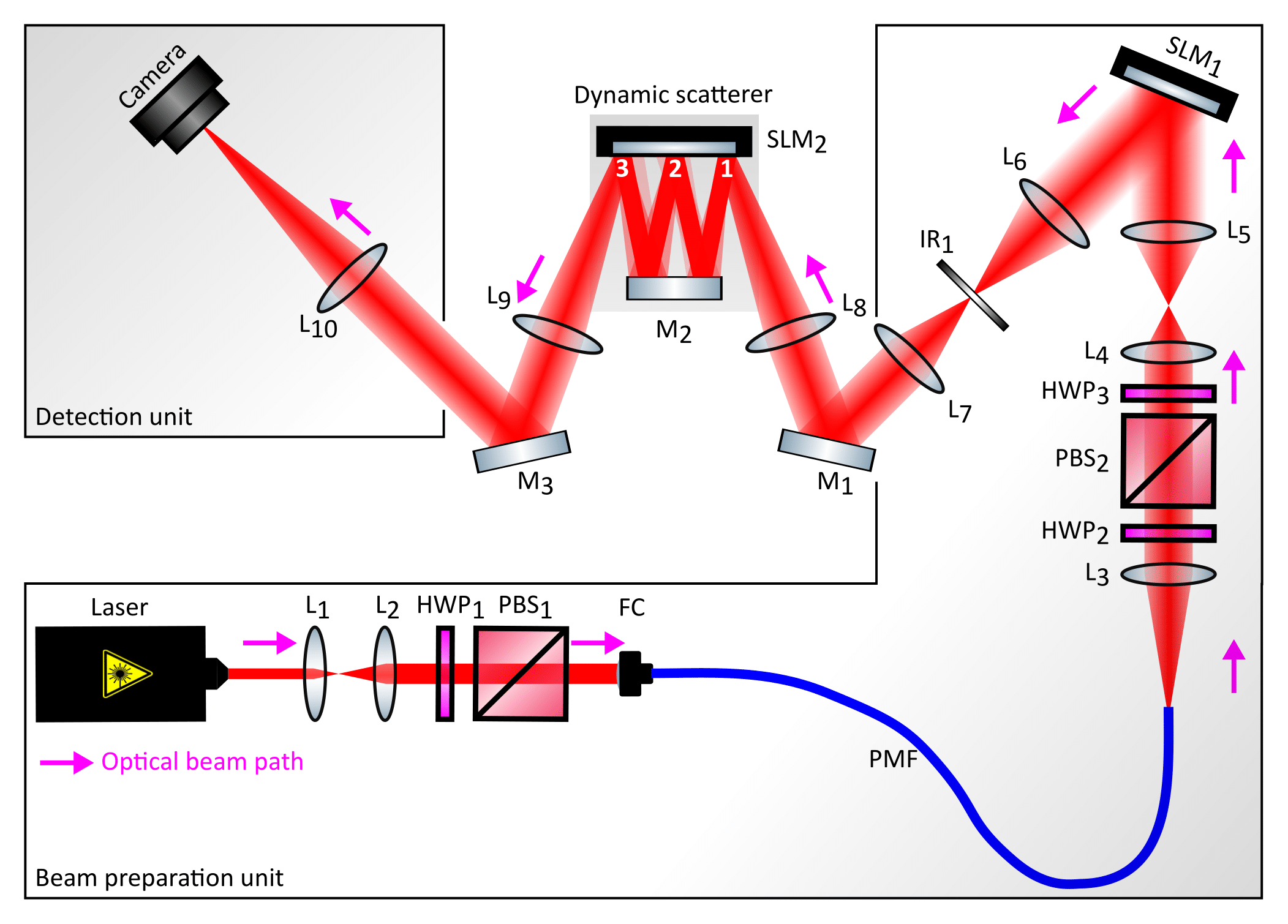}
   \caption{{\bf Experimental setup for unguided optimization}.
   Component information: Laser: Thorlabs-HNL210L. $\text{SLM}_{\text{1}}$: Hamamatsu LCOS 800$\times$600 pixel LC-SLM. $\text{SLM}_{\text{2}}$: Holoeye Pluto-2 1920$\times$1080 pixel LC-SLM. Lenses: $\text{L}_{\text{1}}$ (focal length f\,=\,\SI{50}{\milli\meter}), $\text{L}_{\text{2}}$ (f\,=\,\SI{100}{\milli\meter}), $\text{L}_{\text{3}}$ (f\,=\,\SI{60}{\milli\meter}), $\text{L}_{\text{4}}$ (f\,=\,\SI{150}{\milli\meter}), $\text{L}_{\text{5}}$, $\text{L}_{\text{6}}$ (f\,=\,\SI{300}{\milli\meter}), $\text{L}_{\text{7}}$ (f\,=\,\SI{200}{\milli\meter}),  $\text{L}_{\text{8}}$ (f\,=\,\SI{150}{\milli\meter}), $\text{L}_{\text{9}}$ (f\,=\,\SI{250}{\milli\meter}), $\text{L}_{\text{10}}$ (f\,=\,\SI{200}{\milli\meter})). Mirrors: M$_{1}$, M$_{3}$ are 1" mirrors, M$_{2}$\,=\,10\,mm square mirror (Edmund optics). PMF = polarization maintaining fiber (Thorlabs, P3-630PM-FC-1, length \SI{1}{\meter}). FC = fiber collimator: Thorlabs F110APC-633. HWP = Half wave-plate. PBS = Polarizing beamsplitter. IR = iris. Camera: Basler piA640-210gm. We thank A.\ Frazer for use of the inkscape component library.} 
   \label{Fig:ungiudedSetup}
\end{figure*}

\noindent\S 1: {\bf Unguided optimisation -- experimental details}\\
Figure~\ref{Fig:ungiudedSetup} shows a schematic of the experimental setup used to test our unguided optimisation strategy. In summary: we emulate a dynamic scattering medium using a cascade of 3 phase planes, implemented using 3 reflections from a liquid crystal SLM ($\text{SLM}_{\text{2}}$) with a mirror placed parallel to and facing the SLM chip. The beam incident on the scattering medium is shaped using $\text{SLM}_{\text{1}}$, and a camera records the fluctuations of the output field.

We now describe this setup in more detail: An optical beam emanating from a continuous wave laser (wavelength \SI{632.8}{\nano\meter}, power \SI{21}{\milli\watt}, linearly polarized) is expanded using lenses $\text{L}_{\text{1}}$-$\text{L}_{\text{2}}$ and coupled into a polarization maintaining fiber (PMF) using a fiber collimator (FC). The optical power entering the system is adjusted using a combination of a half wave-plate ($\text{HWP}_{\text{1}}$) and a polarizing beam-splitter ($\text{PBS}_{\text{1}}$). The optical beam diverging from other end of the fiber is collimated using lens $\text{L}_{\text{3}}$. The beam's optical power and polarization is controlled using the combination of half-wave-plates ($\text{HWP}_{\text{2}}$ - $\text{HWP}_{\text{3}}$) and polarizing beam-splitter ($\text{PBS}_{\text{2}}$). The beam is then expanded using lenses $\text{L}_{\text{4}}$-$\text{L}_{\text{5}}$ to overfill the active area of $\text{SLM}_{\text{1}}$ -- used to shape the light incident on the dynamic scattering medium. $\text{SLM}_{\text{1}}$ imparts a spatially varying phase delay to the incident beam, and is also encoded with a phase grating to diffract the desired optical field into the $\text{1}^{\text{st}}$ diffraction order. Light diffracted into other unwanted diffraction orders is blocked out by an iris ($\text{IR}_{\text{1}}$). The spatially filtered beam is re-imaged using lenses $\text{L}_{\text{7}}$-$\text{L}_{\text{8}}$ to the dynamic scattering medium, emulated by $\text{SLM}_{\text{2}}$. The beam sequentially reflects from three separate regions of $\text{SLM}_{\text{2}}$. The outgoing beam is then re-imaged using lenses $\text{L}_{\text{9}}$-$\text{L}_{\text{10}}$ to the camera.

The unguided optimisation proceeds as follows: $\text{SLM}_{\text{1}}$ is subdivided into 1200 equally sized square super-pixels. Initially each superpixel imparts a random phase to incident light. Our aim is to optimise the phase of these super-pixels such that the transmitted field circumnavigate any dynamic regions of the sample, which is emulated using $\text{SLM}_{\text{2}}$.
$\text{SLM}_{\text{2}}$ is set to display a static smoothly varying phase pattern with three square patches (of size $\sim 20\times20$ SLM pixels) inside which the phase dynamically varies. The location of these three patches each coincide with a different reflection area of the light propagating through the SLM-mirror system. $\text{SLM}_{\text{2}}$ updates at a rate of 20\,Hz and cycles through a series of pre-saved patterns as the camera records 100 frames, from which level of fluctuations are calculated. This procedure is repeated for a second test input field, for which 600 randomly chosen super-pixels have their phase uniformly modulated by +pi/40\,rad. Finally, this procedure is repeated for a third test input field, for which the same 600 randomly chosen super-pixels have their phase uniformly modulated by -pi/40\,rad. The incident field which corresponds to the lowest level of fluctuations is used as the starting field for the next iteration of the optimisation. We run our optimisation until the level of fluctuations no longer reduce, which in this case was after about 3000 iterations.\\

\noindent\S 2: {\bf Numerical reconstruction of the fields at each phase plane inside the emulated dynamic scatterer}\\
As we have full control over the test scattering medium, we are able to digitally ‘peel back’ the outer scattering layers to look inside and directly observe the evolution of the optimised field as it propagates through the cascade of phase planes. Experimentally this is achieved by switching-off the aberrating effect of the outer planes, and imaging the optimised field that is incident on plane 2. We recover the phase of this optical field using phase-stepping full-field holography, see e.g.\ ref.~\cite{li2021memory}. The coherent reference needed for this measurement is obtained by splitting off part of the laser beam at $\text{PBS}_{\text{2}}$. $\text{SLM}_{\text{1}}$ is then used to globally phase shift the incident field with respect to the reference, enabling measurement of the field at plane 2 by imaging it along with the coherent reference beam, onto the camera. We measure the field at plane 2 twice: firstly with the phase profile of planes 2 and 3 switched off (i.e.\ set to uniform 0 everywhere) -- yielding the field $\Vecq_1$ and secondly with only the phase profile of plane 3 switched off -- yielding $\Vecq_2$, which includes the effect of the second phase plane. $\Vecq_1$ and $\Vecq_1$ essentially represent the field immediately before and immediately after the second phase plane. Numerically back-propagating $\Vecq_1$ a distance $\delta z$ yields a reconstruction of $\Vecq_0$ -- the optimised field just after plane 1. The numerical propagation is carried out using the angular spectrum method. Numerically forward-propagating $\Vecq_2$ through free-space a distance $\delta z$ yields $\Vecq_3$ -- the optimised field just before plane 3. The fields adjacent to each plane, $\Vecq_0$, $\Vecq_1$ and $\Vecq_3$ are the three fields shown in Fig.~1(e) second row.\\

\noindent\S 3: {\bf Physical adjoint optimisation -- derivation of method}\\
Here we give a full derivation of the physical adjoint optimisation method. Consider a 2D system of randomly arranged scatterers between a source plane and a detector plane. The source field is given by $\textbf{u}(x)$, and the field at the detectors is given by $\textbf{v}(x,t)$. We can then write the field on the detector due to the configuration of particles at time $t$
\begin{equation}
    \textbf{v}_t(x) = \int G_t(x, x') \textbf{u}(x') d^3x,
    \label{eq:1}
\end{equation}
where the integral over the Green's function $G_t$ for the configuration of particles at time $t$ is equivalent to multiplying the transmission matrix $\VecT$ by the source field - the notation used in the main body of this paper.

To minimise the fluctuations in the transmitted field, we aim to maximise the overlap between the output modes at all times $t$, given by $F$:
\begin{equation}
    F=\left|\sum_t \sum_{t'} \int d^3x\ \textbf{v}_t(x)\textbf{v}_{t'}^*(x) \right|^2.
\end{equation}
\noindent In order to understand how to iteratively maximise $F$, we consider a small  change in the output field of $\delta\textbf{v}_t$:
\begin{align}
F&=\left|\sum_t\sum_{t'}\int d^3x\ \left[\textbf{v}_t(x)+\delta\textbf{v}_t\right]\left[\textbf{v}_{t'}^*(x)+\delta\textbf{v}_{t'}^*\right]\right|^2\\
&=\left|\sum_t\sum_{t'}\int d^3x\ \left[\textbf{v}_t(x)\textbf{v}_{t'}^*(x)+\textbf{v}_t(x) \delta\textbf{v}_{t'}^*+\delta\textbf{v}_t\textbf{v}_{t'}^*(x)+\color{red}{\delta\textbf{v}_t\delta\textbf{v}_{t'}^*}\color{black} \right] \right|^2, 
\end{align}
where we drop the red term since it is the product of two small numbers and so can be considered negligible. We write ${z=\delta\textbf{v}_t\textbf{v}_{t'}^*(x)}$ and its complex conjugate $z^*=\textbf{v}_t(x)\delta\textbf{v}_{t'}^*$. Then
\begin{align}
    F&\sim \left|\sum_t\sum_{t'}\int d^3x\ \left[\textbf{v}_t(x)\textbf{v}_{t'}^*(x)+z^*+z \right] \right|^2\\
    &\sim \left|\sum_t\sum_{t'}\int d^3x\ \left[\textbf{v}_t(x)\textbf{v}_{t'}^*(x)+2\text{Re}[z] \right] \right|^2\\
    &\sim \left|\sum_t\sum_{t'}\int d^3x\ \left(\textbf{v}_t(x)\textbf{v}_{t'}^*(x)+2\text{Re}\left[\delta\textbf{v}_t\textbf{v}_{t'}^*(x) \right] \right) \right|^2\\
    &\sim\left|M+2\text{Re}\left[\sum_t\sum_{t'}\int d^3x\ \delta\textbf{v}_t\textbf{v}_{t'}^*(x)\right] \right|^2,\label{eq:small_change}
\end{align}
where we have let $M=\sum_t\sum_{t'}\int d^3x\ \textbf{v}_t(x)\textbf{v}_{t'}^*(x)=\int d^3x\left(\sum_t\textbf{v}_t\right)\cdot\left(\sum_{t'}\textbf{v}_{t'}\right)^{\star}$, which is always positive and real.  In the second term on the right of Eq. (\ref{eq:small_change}) the sum over $t'$ can be moved inside the integral and onto $\textbf{v}_{t'}^*$ as this is the only quantity that depends on $t'$.  The change in our figure of merit then becomes
\begin{equation}
    F\sim\left|M+2\text{Re}\left[\sum_t\int d^3x\ \delta\textbf{v}_t\sum_{t'}\textbf{v}_{t'}^*(x) \right] \right|^2.
\end{equation}
We now rewrite the sum of $\textbf{v}^*_{t'}$ over $t'$ in terms of the time average of the transmitted field,
\begin{equation}
    \sum_{t'}\textbf{v}_{t'}^*\left(x\right)=n\left< \textbf{v}^*\left(x\right)\right>,
\end{equation}
where $n$ is the total number of time points and $\left<\textbf{v}^* \right>$ is the time averaged field.  In terms of this averaged field the figure of merit now equals
\begin{equation}
    F\sim \left|M+2\text{Re}\left[\sum_t\int d^3x\  \delta\textbf{v}_t\ n\left<\textbf{v}^*\left(x\right) \right>\right] \right|^2.
    \label{eq:2}
\end{equation}
At this point we use the relationship between input and output fields (Eqn.~\ref{eq:1}) to write the small change in the transmitted field in terms of a small change in the input field, $\delta \textbf{u}(x')$ (which we note should not depend on time)
\begin{equation}
\delta\textbf{v}_t\left(x\right)=\int d^3x'\ G_t\left(x,x'\right)\delta \textbf{u}(x').
\end{equation}
Applying this representation of the change in the field to Eqn.~\ref{eq:2}, the figure of merit can now be written in terms of the variable $\delta\textbf{u}$ over which we have control
\begin{equation}
    F\sim \left|M+2n\text{Re}\left[\sum_t\int d^3x\int d^3x'G_t\left(x,x'\right)\delta \textbf{u}(x')\left<\textbf{v}^* (x)\right> \right] \right|^2.
\end{equation}
Once the absolute square in this expression is multiplied out, this yields
\begin{align*}
      F&\sim M^2+\color{red}{\left|2n\text{Re}\sum_t\int d^3x\int d^3x'G_t(x,x')\delta \textbf{u}(x')\left<\textbf{v}^*(x)\right> \right|^2}\color{black}\\
    &\ \ \ \ \ \ \ +4nM\text{Re}\left[\sum_t\int d^3x\int d^3x'G_t(x,x')\delta \textbf{u}(x')\left<\textbf{v}^*(x)\right>\right].
\end{align*}
Here, we drop the red term, which is the square of a small number and so negligible compared to the other terms. To increase our figure of merit we thus require
\begin{equation}
    \delta F= 4nM\text{Re}\left[\sum_t\int d^3x\int d^3x'G_t(x,x')\delta \textbf{u}(x')\left<\textbf{v}^*(x)\right>\right]>0\label{eq:dfom}
\end{equation}
As the system is reciprocal $G_{t}(x,x')=G_{t}(x',x)$, Eq. (\ref{eq:dfom}) states that to increase the figure of merit we must propagate the conjugate of the time averaged field back through the time varying system, average this result (the sum over $t$) on the input side, and then choose $\delta\textbf{u}$ so that its overlap with this averaged field has the largest possible real part.
There are several ways of achieving this, but here we modify the input field by a fixed amplitude $\delta A$ and a spatially varying phase signified by column vector $\boldsymbol{\theta}$:
\begin{equation}
    \delta \textbf{u}(x)=\delta A\text{e}^{i\boldsymbol{\theta}},
    \label{eq:dj}
\end{equation}
where we exponentiate each element of the column vector $\boldsymbol{\theta}$ on an element by element basis. Applying Eqn.~(\ref{eq:dj}) to (\ref{eq:dfom}), we can determine this phase $\boldsymbol{\theta}$ via
\begin{align}
    \delta F &= 4nM\text{Re}\left[\delta A\sum_t\int d^3x\, {\rm e}^{i\boldsymbol{\theta}}{\color{blue}\int d^3x'G_t(x,x') \left<\textbf{v}^*(x)\right>}\right].\label{eq:arg_deltaF}
\end{align}
To maximize the value of Eqn.~(\ref{eq:arg_deltaF}) we choose the phase $\boldsymbol{\theta}$ such that it equals the negative of the argument of the term highlighted in blue in Eqn.~\ref{eq:arg_deltaF}
\begin{equation}
    \boldsymbol{\theta}=-\text{arg}\left[\sum_t\int G_t(x',x)\left<\textbf{v}^*(x)\right>d^3x \right]
\end{equation}
With this choice the figure of merit $F$ will be increased by the largest amount, for a fixed small value of $\delta A$. This expression is equivalent to Eqn.~\ref{Eq:conjAv} in the main text. 

We also tested two other ways to iteratively update the input field: Firstly, we found that our optimiser performed similarly if updating the input field according to 
\begin{equation}
\delta\Vecu=N\left[\sum_t\int G_t(x',x)\left<\textbf{v}^*(x)\right>d^3x \right]^*,
\end{equation}
where $N$ is a normalisation constant -- i.e.\ including both phase and amplitude information from the time-reversed average field sent back through the dynamic scatterer. Secondly, we also found similar results by simply replacing the input field with the time-reversed average field emerging onto the left hand side of the scatterer. This final option essentially making the optimisation process symmetrical about the scattering medium. All of these methods work since the updated field appears to become closer to a stable field that avoids moving parts of the medium after each iteration, although we did not conduct a detailed analysis of the rate of convergence for the different methods.
\\

\begin{figure*}[p]
  \includegraphics[width=0.9\textwidth]{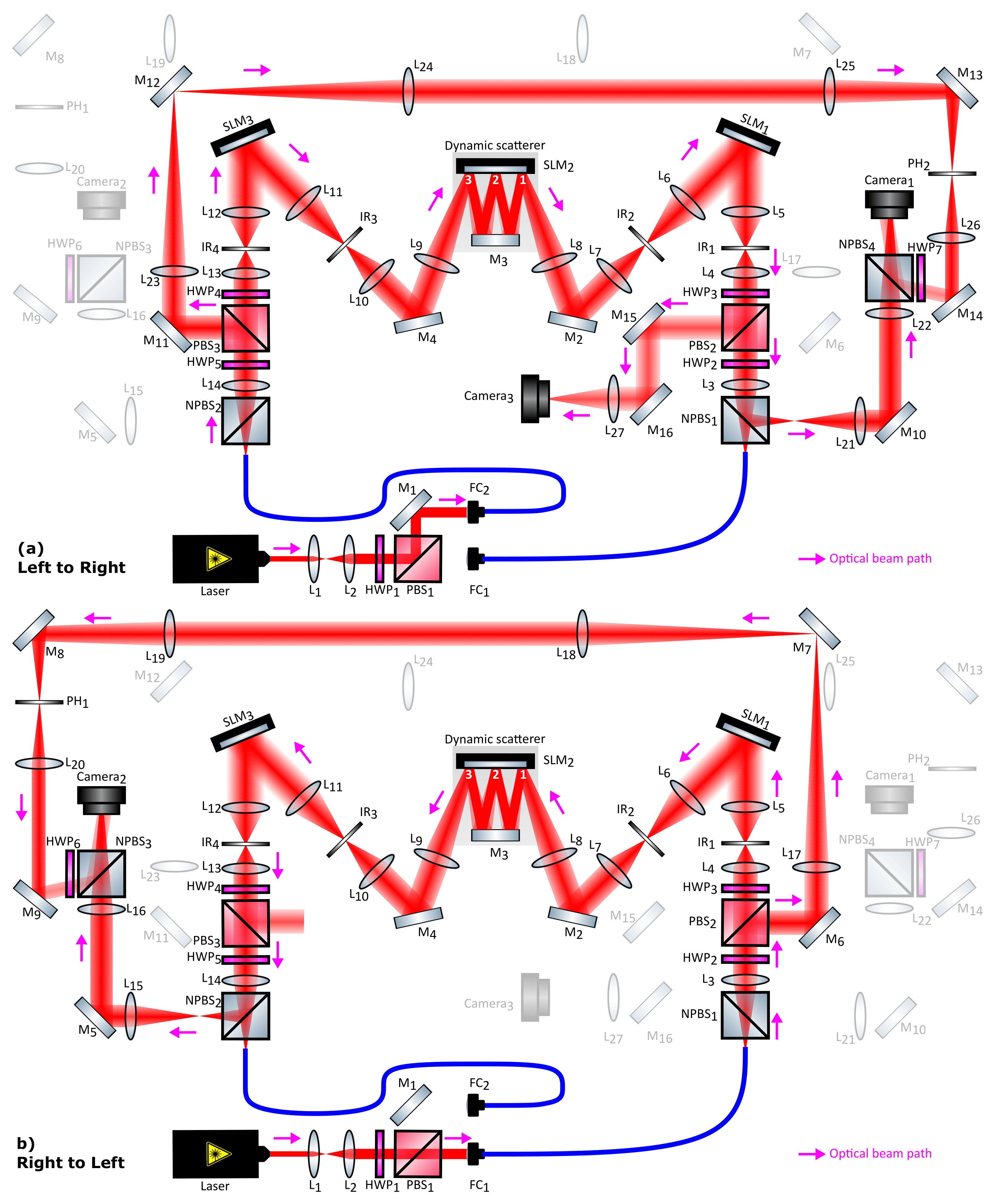}
   \caption{{\bf Experimental setup for physical adjoint optimization}. (a) Configuration for light to be sent through the scattering medium from left to right. (b) Configuration for light to be sent through the scattering medium from right to left. Description of the components: Laser: Thorlabs-HNL210L. SLM$_1$ and SLM$_2$: Holoeye Pluto 2 LC-SLM. SLM$_3$: Vialux V-7001 DMD. Lenses: $\text{L}_{\text{1}}$ (f\,=\,\SI{50}{\milli\meter}), $\text{L}_{\text{2}}$ (f\,=\,\SI{100}{\milli\meter}), $\text{L}_{\text{3}}$, (f\,=\,\SI{60}{\milli\meter}), $\text{L}_{\text{4}}$, (f\,=\,\SI{150}{\milli\meter}),  $\text{L}_{\text{5}}$, $\text{L}_{\text{6}}$ (f\,=\,\SI{300}{\milli\meter}), $\text{L}_{\text{7}}$ (f\,=\,\SI{200}{\milli\meter}), $\text{L}_{\text{8}}$ (f\,=\,\SI{150}{\milli\meter}), $\text{L}_{{\text{9}}}$ (f\,=\,\SI{150}{\milli\meter}), $\text{L}_{\text{10}}$ (f\,=\,\SI{200}{\milli\meter}),  $\text{L}_{\text{11}}$ (f\,=\,\SI{300}{\milli\meter}), $\text{L}_{\text{12}}$ (f\,=\,\SI{300}{\milli\meter}), $\text{L}_{\text{13}}$ (f=\SI{150}{\milli\meter}), $\text{L}_{\text{14}}$ (f\,=\,\SI{60}{\milli\meter}) , $\text{L}_{\text{15}}$ (f\,=\,\SI{35}{\milli\meter}), $\text{L}_{\text{16}}$ (f\,=\,\SI{200}{\milli\meter})), $\text{L}_{\text{17}}$ (f=\SI{300}{\milli\meter}) and $\text{L}_{\text{18}}$ (f\,=\,\SI{300}{\milli\meter}); $\text{L}_{\text{19}}$ and $\text{L}_{\text{20}}$. PMF = polarization maintaining fiber: Thorlabs, P3-630PM-FC-1, length \SI{1}{\meter}. HWP = Half wave-plate. PBS = polarizing beam splitter. NPBS = Non-polarizing beam splitter. IR = iris. PH = Pinhole. Cameras 1,2: basler piA640-210gm, 640\,$\times$\,480 pixels. Camera 3: Thorlabs Thorcam DCC1545M-GL.}
   \label{Fig:Adjoint}
\end{figure*}

\noindent\S 4: {\bf Physical adjoint optimisation -- experimental details}\\
Figure~\ref{Fig:Adjoint} shows a schematic of the experimental setup used to test the physical adjoint optimisation strategy. The dynamic scattering medium is created in the same way as described above, using 3 reflections from an SLM.

The dynamic medium is first illuminated from the left hand side, as shown in the upper schematic in Fig.~\ref{Fig:Adjoint}. The laser (same as detailed above) is expanded and half-wave-plate HWP$_1$ is set so that light reflects at beam-splitter PBS$_1$. The beam is split into a signal and reference arm using polarising beam-splitter PBS$_3$. The signal beam is shaped by SLM$_3$, which in our experiment is a digital micro-mirror device (DMD). Here we implement intensity and phase beam shaping using the DMD as described in~\cite{lee1979binary}, and iris IR$_3$ blocks the unwanted diffraction orders. The shaped beam passes through the dynamic medium (SLM$_2$) to camera 1. The reference beam is directed around the dynamic scatterer to camera 1, and the interference of the two beams enables the time-evolving scattered field to be measured using single-shot off-axis digital holography. The time-averaged optical field is calculated from 5 optical fields -- this number chosen empirically after testing different averaging times. During this phase of the experiment, SLM$_1$ (an LC-SLM) performs no beam shaping, but displays a uniform grating to direct light through iris IR$_1$. 

Light is next sent through the system in the opposite direction, as shown in the lower schematic in Fig.~\ref{Fig:Adjoint}. To switch the laser direction, half-wave-plate HWP$_1$ is rotated so light transmits through beam-splitter PBS$_1$. If necessary, this step could be automated for fast operation in the future. The beam is then split into a signal and reference arm using polarising beam splitter PBS$_2$. The signal beam is now shaped by SLM$_1$, to create the phase conjugate of the time-averaged optical field measured in the previous step. Here we implement intensity and phase beam shaping using the phase-only SLM as described in~\cite{davis1999encoding}, and iris IR$_2$ blocks the unwanted diffraction orders. The beam passes through the dynamic medium (SLM$_2$) to camera 2. The reference beam is directed around the dynamic scatterer to camera 2, and the interference of the two beams once again enables the time-evolving scattered field to be measured using single-shot off-axis digital holography. The time-averaged optical field, and consequently the updated input field, are calculated. During this phase of the experiment, SLM$_3$ performs no beam shaping, but displays a uniform grating to direct light through iris IR$_4$. This completes one iteration of our physical adjoint optimisation strategy, which is then repeated until the optimised field converges. We note that the correct alignment of the digital optical phase conjugation systems on either side of the sample is critical to the performance of the system. Our alignment procedure is explained in detail in ref.~\cite{mididoddi2020high}. Once the optimisation is complete, Camera 3 is used to observe the propagation of the optimised beam inside the medium.\\

\begin{figure*}[h]
  \includegraphics[width=0.8\textwidth]{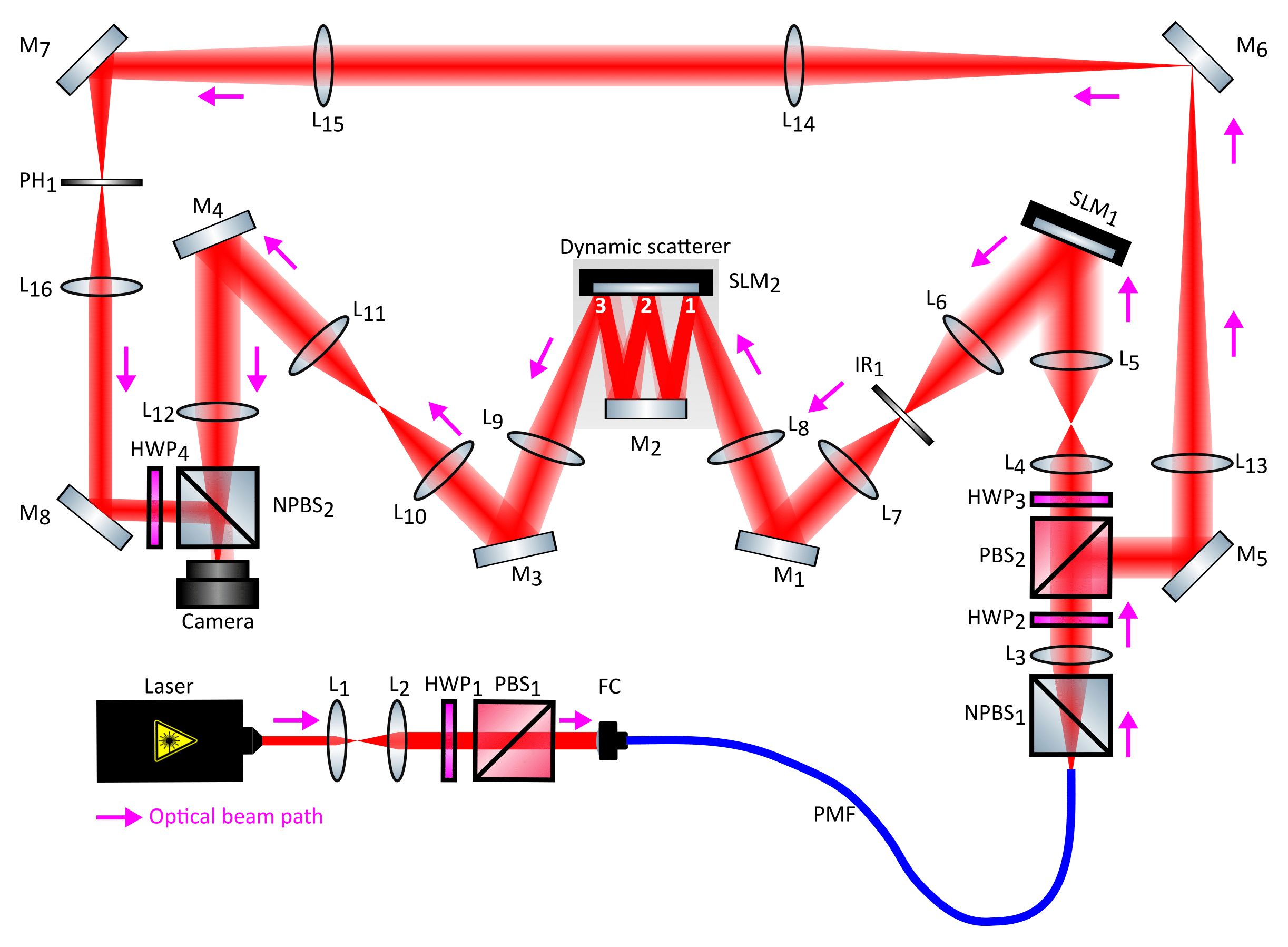}
   \caption{{\bf Experimental setup for time-averaged transmission matrix measurement}.
   Description of the components: Laser: Thorlabs-HNL210L. SLM$_1$, SLM$_2$: Holoeye Pluto 2 LC-SLM. Lenses: $\text{L}_{\text{1}}$ (f\,=\,\SI{50}{\milli\meter}), $\text{L}_{\text{2}}$ (f\,=\,\SI{100}{\milli\meter}), $\text{L}_{\text{3}}$, (f\,=\,\SI{60}{\milli\meter}), $\text{L}_{\text{4}}$, (f\,=\,\SI{150}{\milli\meter}),  $\text{L}_{\text{5}}$, $\text{L}_{\text{6}}$ (f\,=\,\SI{300}{\milli\meter}), $\text{L}_{\text{7}}$ (f\,=\,\SI{200}{\milli\meter}), $\text{L}_{\text{8}}$ (f\,=\,\SI{150}{\milli\meter}), $\text{L}_{{\text{9}}}$ (f\,=\,\SI{150}{\milli\meter}), $\text{L}_{\text{10}}$ (f\,=\,\SI{200}{\milli\meter}), $\text{L}_{\text{11}}$ (f=\SI{75}{\milli\meter}), $\text{L}_{\text{12}}$ (f=\SI{175}{\milli\meter}), $\text{L}_{\text{13}}$ (f=\SI{300}{\milli\meter}), $\text{L}_{\text{14}}$ (f\,=\,\SI{300}{\milli\meter}), $\text{L}_{\text{15}}$ (f\,=\,\SI{100}{\milli\meter}), $\text{L}_{\text{16}}$ (f\,=\,\SI{50}{\milli\meter})). PMF = polarization maintaining fiber: Thorlabs, P3-630PM-FC-1, length \SI{1}{\meter}). HWP = Half wave-plate. PBS = polarizing beam splitter. NPBS = Non-polarizing beam splitter. IR = iris. PH = Pinhole. Camera: basler piA640-210gm, 640\,$\times$\,480 pixels.}
   \label{Fig:time-avTM}
\end{figure*}

\noindent\S 5: {\bf Time-averaged transmission matrix -- experimental details}\\
Figure~\ref{Fig:time-avTM} shows a schematic of the experimental setup used to measure the time-averaged TM. The laser beam is split into a signal and reference path at PBS$_2$. The light in the signal beam path is shaped by SLM$_1$, and passes through the dynamic scattering medium to the camera, where it interferes with light from the reference beam path. A sequence of 2304 orthogonal input probe fields are transmitted through the medium, and the output field is measured using single-shot off-axis digital holography. The output field is averaged over 10 configurations of the dynamic medium for each probe mode. It is critical that the phase of the reference beam does not drift with respect to the light in the signal arm throughout this measurement. A standard approach to negate the effect of phase drift during TM measurement is to interlace the changing input probe measurements with the projection of a standard probe field, the mode of which is held constant throughout the measurement. The global phase of this probe field tracks the phase drift between the arms of the interferometer, which can then be subtracted. Here, we found that this method can still be applied to a partially moving sample -- as long as the sample fluctuations are minor, the global phase of a standard probe mode faithfully tracks the phase drift. Ideally the probe mode would itself be stable, and so in some cases it may be necessary to first optimise a single stable probe mode to use for drift tracking, before measuring the time-averaged TM. In our experiments we found such a pre-optimisation step to be unnecessary.\\

\noindent\S6: {\bf Focusing through dynamic scattering media using the time-averaged transmission matrix}\\
In order to use the information held in the time-averaged TM for stable beam shaping through dynamic scattering media we first calculate the eigenvectors of matrix ${\VecT_{\text{av}}^\dagger\VecT_{\text{av}}}$, yielding: ${\VecT_{\text{av}}^\dagger\VecT_{\text{av}} = \VecU\Lambda\VecU^\dagger}$, where as is convention, $\VecU$ is a matrix of eigenvectors held on its columns, and $\Lambda$ a diagonal matrix of eigenvalues. We build a new matrix $\VecU'$ that retains only the columns of $\VecU$ that have associated eigenvalues with absolute values greater than a stipulated threshold level. The number of eigenvalues above the threshold counts the number of independent fields that are able to thread around any moving regions of the medium. Therefore we can recover a static transmission matrix $\VecV = \VecT\VecU'$ and use it to generate a target output field $\Vecv_{\text{trg}}$ by injecting $\Vecu = \VecU'\VecV^\dagger\Vecv_{\text{trg}}$ (where here we have taken the conjugate transpose as an approximation to the inverse of $\VecV$). Here the input field $\Vecu$ and output field $\Vecv_{\text{trg}}$ are represented in the input and output bases originally used to measure the time-averaged TM. Note that when defining matrix $\VecV$, then matrix $\VecT$ can be any of the previously measured TMs $\VecT(t)$, or the average TM $\VecT_{\text{av}}$, since the fields described by $\VecV$ should not impinge on the moving regions of the complex medium.

\noindent\S7: {\bf Description of supplementary movies}\\
\noindent {\bf Movie 1}: Camera frames showing the time-dependent intensity profile of light transmitted through a dynamic scattering medium. Left panel: initial field. Right panel: Optimised field, found using unguided optimisation. We observe that the optimised field exhibits much lower levels of time-dependent fluctuations than the initial field. Scale bar shows relative intensity.\\

\noindent {\bf Movie 2}: Camera frames showing the time-dependent intensity profile of light transmitted through a dynamic scattering medium. Left panel: initial field. Right panel: Optimised field, found using physical adjoint optimisation. We observe that the optimised field exhibits much lower levels of time-dependent fluctuations than the initial field. Scale bar shows relative intensity.\\

\noindent {\bf Movie 3}: Camera frames showing the time-dependent intensity profile of light transmitted through a dynamic scattering medium. Each panel shows the transmitted light when a particular fluctuation eigenchannel is excited (the eigenchannel index is labelled above each panel). We observe that the light transmitted through the high index eigenchannels fluctuates the least, while light transmitted through low index eigenchannels exhibits much higher levels of temporal fluctuations. Scale bar shows relative intensity.\\

\noindent {\bf Movie 4}: Camera frames showing the time-dependent intensity profile of light transmitted through a dynamic scattering medium. Left panel: the result of attempting to form a focus at the output using the conventional inverse TM. Right panel: the focus generated through the same sample, formed by exciting only the top 100 most stable fluctuation eigenchannels. In both cases we scan the position of the focus over 4 different positions. We observe that the focus created using the stable eigenchannels is enhanced: it haS a higher contrast, and exhibits lower levels of temporal fluctuations. Scale bar shows relative intensity.\\

\end{document}